  \documentstyle[aps,multicol]{revtex}

\draft
\title{Magnetoinductance of Josephson junction array \\
with frozen vortex diffusion}
\author{S. E. Korshunov}
\address{L. D. Landau Institute for Theoretical Physics,
         Kosygina 2, 119334 Moscow, Russia}
\date{March 27, 2003}

\begin{document}
\maketitle

\begin{abstract}
The dependence of sheet impedance of a Josephson junction array
on the applied magnetic field is investigated in the regime when
vortex diffusion between array's plaquettes is
effectively frozen due to low enough temperature.
The field dependent correction to sheet inductance is found to
be proportional to $f\ln(1/f)$, where $f\ll 1$ is the magnitude of the
field expressed in terms of flux quanta per plaquette.
\end{abstract}
\pacs{PACS numbers: 74.81.Fa, 74.25.Nf}

 \begin{multicols}{2}
\section{Introduction}

A two-coil mutual-inductance technique \cite{FH,JGRLM}
has been introduced by Fiory and Hebard \cite{FH} for measurement
of (linear) frequency dependent complex sheet impedance
$Z_\Box(\omega)$ [or, equivalently, sheet conductance
$G_\Box(\omega)=1/Z_\Box(\omega)$]
of thin superconducting films \cite{HF}.
This approach has also proved itself very useful for investigation
of arrays of weakly coupled superconducting islands (Josephson
junction arrays) in external magnetic field (for a review,
see Ref. \onlinecite{ML}).
In such systems the nature of the response is strongly dependent
on the magnitude of the applied dc field,
which is convenient to discuss in terms of the ratio $f=\Phi/\Phi_0$
of the magnetic flux $\Phi$ penetrating each array plaquette
to the superconducting flux quantum $\Phi_0$.

For an integer (for example, zero) $f$ a Josephson junction array can
be described by the two-dimensional $XY$ model and with decrease of
temperature experiences the Berezinskii-Kosterlitz-Thouless transition
\cite{Ber,KT} into a superconducting phase. In this phase all vortices
are bound in neutral pairs and the sheet impedance in the limit of low
frequency becomes purely inductive: $Z_\Box(\omega)\approx i\omega
L_\Box$, the effective sheet inductance $L_\Box$ being inversely
proportional to the superfluid density. A deviation of $f$ from an
integer value (by $\delta f$) introduces a finite concentration of
unbound vortices $c$ (in dimensionless units, that is per lattice
plaquette, $c=|\delta f|$), as a consequence of which the response is
strongly changed and in the low frequency limit becomes purely
dissipative: $Z_\Box(\omega)\approx cR_{\rm V}$ (the frequency
independent constant $R_{\rm V}$ being associated
with a contribution from a single vortex).

Analogous behavior can be expected to take place in the vicinity of
every rational $f$, if the temperature is lower than the
(discontinuously dependent on $f$) temperature of the phase transition
associated with freezing of the field-induced vortices into
commensurate pattern \cite{TJ}.
As a consequence, a measurement of the real and
of the imaginary components of $Z_\Box(\omega)$ as a function of $f$
(at fixed $\omega$ and reasonably chosen temperature)
produces strongly oscillating curves \cite{T,T1,T2,T3,M}, the well
developed dips on which correspond to the superconducting states.

With further decrease of temperature the dissipative component of
$Z_\Box(\omega)$ becomes very small and the oscillations of the
inductive component much less pronounced \cite{T,T1,M}. This
is related with suppression of vortex diffusion at temperatures
for which the rate of a thermally activated tunnelling of a vortex
between neighboring lattice plaquettes (which requires to overcome
a well defined barrier \cite{LAT})
becomes much smaller than the frequency of the measurement. 

The present work is devoted to investigation of a Josephson junction
array response in this particular regime,
when the relation between the frequency and the temperature allows
to consider all vortices as frozen in lattice cells which they occupy.
In terms of a simple model \cite{ML,LAT} treating a vortex
in a proximity junction array as an overdamped point particle
moving in the external potential (imposed by the structure of the
array) this corresponds to the case when each of the vortices is
confined to oscillate within the limits of a particular minimum of the
potential, which then can be replaced by a harmonic one. In the
framework of such description the correction to sheet inductance (per
vortex) would have a finite value, inversely proportional to the
curvature of the potential.

With the help of a more straightforward calculation (based on
the reduction to the equivalent electric circuit) we show that the
single vortex contribution contains logarithmic divergence and,
therefore, for small magnetic fields (that is small vortex
concentration, $f\ll 1$) the correction to $L_\Box^{-1}$ is
proportional to $f\ln(1/f)$. The results can be of interest in
relation with sheet impedance measurements for small $f$ at
intermediate frequencies \cite{T1,T2}.

\section{The Hamiltonian and the vortices}

In absence of external magnetic field
a regular Josephson junction array can be described by the Hamiltonian
\begin{equation}
H=-J\sum_{({\bf n}{\bf n'})}\cos(\varphi_{\bf n}-\varphi_{\bf n'})
\;,                                                        \label{H}
\end{equation}
where $\varphi_{\bf n}$ is the phase of the order parameter of
${\bf n}$-th superconducting island.
In the case of a square lattice the variable ${\bf n}$ can
be chosen in the form of the vector ${\bf n}=(n_x,n_y)$ with integer
components $n_x$ and $n_y$.
The summation in Eq. (\ref{H}) is performed over all pairs
$({\bf nn'})$ of coupled islands and
\begin{equation}
J=\frac{\hbar}{2e}I_c
\label{J} \end{equation}
is the Josephson coupling constant, which is assumed to be the same
for all junctions, $I_c$ being the critical current of a single
junction. The form of Eq. (\ref{H}) assumes that the coupling is weak
enough, so the magnetic field created by the currents flowing in
the array can be neglected.

Variation of the Hamiltonian (\ref{H}) with respect to
$\varphi_{\bf n}$ shows that the minimums of $H$ are achieved when
the variables $\{\varphi_{\bf n}\}$ satisfy the current conservation
equations
\begin{equation}
\sum_{{\bf n'}}I_{\bf nn'}=0 \;,                          \label{CC}
\end{equation}
where
\begin{equation}
I_{\bf nn'}=I_c\sin(\varphi_{\bf n'}-\varphi_{\bf n})     \label{C}
\end{equation}
is the current from the ${\bf n}$-th to the ${\bf n'}$-th island,
defined only for the pairs of the islands connected by the junction.
The simplest solution of the Eqs. (\ref{CC})-(\ref{C}) is the trivial
solution
\begin{equation}
\varphi_{\bf n}=\mbox{const}
\end{equation}
corresponding to the global minimum of $H$ and the absence of any
currents.

A vortex is a local minimum of $H$, in which on going along any
closed loop surrounding the vortex core (which can be associated with
a particular plaquette of the lattice) $\varphi_{\bf n}$ changes by
$2\pi s$, where $s=\pm 1$ is the topological charge of a vortex.
The form of this solution implies the presence of persistent currents
circulating around the vortex core.

Away from the core $\varphi_{\bf n}$ changes slowly and Eqs.
(\ref{CC})-(\ref{C}) can be linearized to give:
\begin{equation}
\sum_{{\bf n'}}(\varphi_{\bf n'}-\varphi_{\bf n})=0\;,  \label{llap}
\end{equation}
where the summation [like in Eq. (\ref{CC})] is performed over the
nearest neighbors of ${\bf n}$.
In continuous approximation Eq. (\ref{llap}) is reduced to
\begin{equation}
\nabla^2\varphi=0\;,                                  \label{lapl}
\end{equation}
which allows to conclude that at large distances from the vortex core
the spatial distribution of $\varphi$ is given by
\begin{equation}
\varphi^{v}(x,y)\approx s\arctan\frac{y}{x}+\mbox{const}\;,
                                                    \label{Phi}
\end{equation}
where $x$ and $y$ ($r^2\equiv x^2+y^2\gg 1$) are the coordinates
(in lattice units) counted from the core.

Continuous approximation can be also used for estimating
the energy of a vortex
\begin{equation}
E_{v}=J\sum_{{\bf nn'})} [1-\cos (\varphi^{v}_{\bf n}
- \varphi^{v}_{\bf n'}) ] \;,                   \label{EVL}
\end{equation}
because the integral
\begin{equation}
E_{v}\approx\frac{\gamma J}{2}
         \int\int dx\,dy(\nabla\varphi^{v})^2\;, \label{EV}
\end{equation}
to which the lattice sum of Eq. (\ref{EVL}) is reduced in the
framework of this approximation, diverges at the upper limit:
\begin{equation}
E_{v}\approx\pi\gamma J\ln N\;.                     \label{EV2}
\end{equation}
Here $N$ is the linear size of the array and $\gamma$ is the
numerical constant depending on the structure of the lattice
(for square lattice $\gamma=1$ and for triangular one
$\gamma=\sqrt{3}$).
Accordingly, the main contribution to Eq. (\ref{EVL}) is coming
from the large scales, where it can safely be replaced by its
continuous form (\ref{EV}).
As a consequence of this divergence, at low temperatures all vortices,
which appear as thermal fluctuations, are bound in neutral pairs.
On the other hand,
a finite concentration of vortices of the same sign can be induced by
application of external magnetic field in perpendicular to array.

\section{The equivalent electric network}

Expansion of the Hamiltonian (\ref{H}) up to the second order in
deviations
\begin{equation}
\delta\varphi_{\bf n}=\varphi_{\bf n}-\varphi_{\bf n'}^{(0)}
                                                   \label{psi}
\end{equation}
of the variables $\varphi_{\bf n}$ from their values
$\varphi_{\bf n}^{(0)}$ in some of the minima of $H$ gives
\begin{eqnarray}
H^{(2)}\{\delta\varphi\} & = & H\{\varphi^{(0)}+\delta\varphi\}
-H\{\varphi^{(0)}\} \nonumber\\
& \approx & \frac{1}{2}\sum_{\bf nn'}{J_{\bf nn'}}{}
(\delta\varphi_{\bf n}-\delta\varphi_{\bf n'})^2\;,  \label{H2}
\end{eqnarray}
where
\begin{equation}
J_{\bf nn'}=J\cos[\varphi_{\bf n}^{(0)}-\varphi_{\bf n'}^{(0)}]\;.
                                                   \label{Jnn}
\end{equation}
Since the deviation of the current $\delta I_{{\bf nn}'}$
from its value in the extremal solution
$I_{{\bf nn}'}=I_0\sin[\varphi_{\bf n}^{(0)}-\varphi_{\bf n'}^{(0)}]$
in linear approximation is given by
\begin{equation}
\delta I_{{\bf nn}'}=I_c\cos[\varphi_{\bf n}^{(0)}-\varphi_{\bf
n'}^{(0)}]
(\delta \varphi_{\bf n}-\delta \varphi_{\bf n'}) \;,     
\end{equation}
Eq. (\ref{H2}) can be rewritten as
\begin{equation}
H^{(2)}=\frac{1}{2}\sum_{({\bf nn}')}L_{{\bf nn}'}
(\delta I_{{\bf nn}'})^2 \;,                        \label{H2a}
\end{equation}
where
\begin{equation}
L_{\bf nn'}
=\frac{L_0} {\cos[\varphi_{\bf n}^{(0)}- \varphi_{\bf n'}^{(0)}]}
                                                      \label{Lnn}
\end{equation}
and
\begin{equation}
L_0=\left[\frac{\hbar}{2e}\right]^2 \frac{1}{J}
   =\frac{\hbar}{2eI_c}\;.      \label{L0}
\end{equation}

The form of the Hamiltonian (\ref{H2a}) allows to conclude
\cite{YS,KMM} that
in the harmonic approximation the array behaves itself
with respect to additional (external) current
as the network formed by the inductances
$L_{{\bf nn}'}\equiv L_{{\bf n}'{\bf n}}$ defined by Eqs. (\ref{Lnn}).
At finite frequencies it is also necessary to take into account that
each of these inductances is shunted by the resistance $R_{\bf nn'}$,
so the complex conductance $G_{{\bf nn}'}(\omega)$ of each network
link is given by
\begin{equation}
G_{\bf nn'}(\omega)=\frac{1}{i\omega L_{\bf nn'}}
                   +\frac{1}{R_{\bf nn'}} \;.          
\end{equation}
In the case of a proximity coupled array, which is explicitly
considered below,
the shunting resistance $R_{\bf nn'}$ is determined mainly by the
conductivity of the underlying metallic substrate, and, therefore, can
be assumed to be frequency independent and the same for all the links:
$R_{\bf nn'}\equiv R_0$. In absence of vortices $\varphi_{\bf
n}=\mbox{const}$, and, therefore, the conductance of all the links is
the same:
\begin{equation}
G_{\bf nn'}(\omega)=G_0(\omega)
\equiv \frac{1}{i\omega L_0}+\frac{1}{R_0} \;. 
\end{equation}

The distribution of the currents $I_{\bf nn'}(\omega)$ in such network
has to be found by solving the current conservation equations of
the form (\ref{CC}) with
\begin{equation}
I_{\bf nn'}(\omega)=G_{\bf nn'}(\omega)(V_{\bf n}-V_{\bf n'}) \;,
                                                          \label{C2}
\end{equation}
where $V_{\bf n}$ is the amplitude of the time dependent electric
potential
\begin{equation}
V_{\bf n}(t)=V_{\bf n}\exp(i\omega t)
\end{equation}
on the ${\bf n}$-th superconducting island.
Application of the time dependent potential difference
\begin{equation}
V(t)=V\exp(i\omega t)                
\end{equation}
for example in the $x$ direction to the square network $N\times N$
formed by equivalent elements, leads to the distribution of
the potentials
\begin{equation}
V_{\bf n}=-\frac{n_x}{N}V+\mbox{const}                
\end{equation}
and the currents:
\begin{equation}
I_{{\bf nn'}}=\left\{\begin{array}{ll}
G_0(\omega)V/N & \mbox{ for } {\bf n'}={\bf n}+{\bf e}_x \\
0 & \mbox{ for } {\bf n'}={\bf n}+{\bf e}_y \;,
\end{array}\right.                                     
\end{equation}
[where ${\bf e}_x=(1,0)$ and ${\bf e}_y=(0,1)$]
corresponding to the total current in the chosen direction given by
\begin{equation}
I(\omega)=G_0(\omega)V \;.                              \label{I0}
\end{equation}
This means that the sheet conductance $G^{}_\Box(\omega)$ of a uniform
square network coincides with the conductance of a single link:
\begin{equation}
G^{}_\Box(\omega)=G_0(\omega) \;.                     \label{GBox}
\end{equation}
In case of a uniform triangular network
$G_\Box(\omega)=\sqrt{3}\,G_0(\omega)$ and, like in the case of a
square network, the sheet conductance does not depend
on the direction.

In the following it will be convenient to decompose $G_\Box(\omega)$
into the real and the imaginary parts as
\begin{equation}
G_\Box(\omega)\equiv \frac{1}{i\omega L_\Box(\omega)}
                    +\frac{1}{R_\Box(\omega)} \;\;,    
\end{equation}
where the effective sheet inductance $L_\Box(\omega)$ and
the effective shunting resistance $R_\Box(\omega)$ are the real
functions of $\omega$. In particular,  Eq. (\ref{GBox}) corresponds
to $L_\Box(\omega)=L_0$ and $R_\Box(\omega)=R_0$.

\section{The single vortex correction to conductance}

In order to find the correction
to frequency dependent sheet conductance related with the
presence of a vortex (which is assumed to be frozen in a particular
array plaquette) one has to consider a network in which the
distribution of $\varphi_{\bf n}$ and, therefore, of
$G_{\bf nn'}$ is assumed to correspond to vortex configuration.
For non-uniform $\varphi_{\bf n}$
it is convenient to rewrite the expression for the total
current (in the $x$ direction) in a square network as
\begin{eqnarray}
I(\omega) & = & \frac{1}{N}\sum_{\bf n}I_{{\bf n,n+e}_x}(\omega)
\nonumber\\ & = & \frac{1}{N^2}\left[\sum_{\bf n}G_{{\bf n,n+e}_x}
-\sum_{\bf n}v_{\bf n}P_{\bf n}\right]V \;,          \label{C3}
\end{eqnarray}
where
\begin{equation}
P_{\bf n}\equiv G_{{\bf n,n+e}_x}-G_{{\bf n},{\bf n-e}_x}
=\frac{1}{i\omega}\left({L^{-1}_{{\bf n},{\bf n+e}_x}}-
{L^{-1}_{{\bf n},{\bf n-e}_x}}\right)                \label{P}
\end{equation}
and $v_{\bf n}$ parametrizes the deviation of $V_{\bf n}$ from its
value in a uniform network:
\begin{equation}
V_{\bf n}=-\frac{n_x+v_{\bf n}}{N}V+\mbox{const} \;,    \label{vn}
\end{equation}
and has to be found by solving the current conservation equations,
which for $V_{\bf n}$ of the form (\ref{vn}) can be rewritten as
\begin{equation}
\sum_{\bf n'} G_{\bf nn'}(\omega)(v_{\bf n}-v_{\bf n'})
=P_{\bf n} \;.                                          \label{eqv}
\end{equation}

Comparison of Eq. (\ref{C3}) with Eq. (\ref{I0}) shows that the
correction to $G_\Box(\omega)$ induced by
a non-uniform distribution of $\varphi_{\bf n}$ can
be split into two contributions, the first of which has the form of a
frequency independent correction to $L^{-1}_\Box$:
\begin{equation}
(\Delta L^{-1}_\Box)_1=-\frac{1}{N^2}
\sum_{\bf n}(L^{-1}_0-L^{-1}_{{\bf n},{\bf n+e}_x})
=-\frac{L_0^{-1}}{N^2}\frac{E_{\rm }}{2J}              \label{DL1}
\end{equation}
and turns out to be proportional to the energy $E$
(counted from the ground state energy)
of the considered configuration.
In the case of a single vortex this energy is given by Eq. (\ref{EV2})
and, accordingly, the expression for $(\Delta L^{-1}_\Box)_1$
contains the logarithmic divergence.

The form of this correction corresponds to formal averaging
of the superfluid density. In conjunction with the absence of a
frequency dependence this allows to
conclude that $(\Delta L^{-1}_\Box)_1$ has nothing to do with vortex
oscillations in a potential minimum and should be associated with
suppression of superfluid density.

The second correction
\begin{equation}
(\Delta G_\Box)_2=-\frac{1}{N^2}\sum_{\bf n}v_{\bf n}P_{\bf n}
                                                       \label{DL2}
\end{equation}
is characterized by a more complex frequency dependence and reduces
to correction to
$L^{-1}_\Box$ only in the limit of $\omega\rightarrow 0$.
It is not hard to show that in the case of a single vortex
the lattice sum in Eq. (\ref{DL2})
[in contrast to the lattice sum in Eq. (\ref{DL1})]
is not divergent at large scales.

The behavior of $P_{\bf n}$ away from the vortex core can be found
from the continuous approximation, in the framework of which
\begin{equation}
P_{\bf n}
\approx \frac{1}{i\omega L_0}\frac{\partial}{\partial x}
\cos\left(\frac{\partial \varphi}{\partial x}\right)
\approx -\frac{1}{2i\omega L_0}\frac{\partial}{\partial x}
\left(\frac{\partial\varphi}{\partial x}\right)^2 \;\;. \label{P1}
\end{equation}
For $\varphi(x,y)$ of the form (\ref{Phi}) this gives
\begin{equation}
P_{\bf n}\approx\frac{2}{i\omega L_0}\frac{xy^2}{(x^2+y^2)^3}
                                             \;. \label{P2}
\end{equation}
Therefore, if $v_{\bf n}$ decays with the increase of the distance
from the vortex core (as naturally one can expect it to do), the
lattice sum in the right-hand side of Eq. (\ref{DL2}) will be
convergent and the behavior of correction to $L^{-1}_\Box(\omega=0)$
will be dominated by the divergence of the contribution
$(\Delta L^{-1}_\Box)_1$ discussed above.

The behavior of $v_{\bf n}$ away from the vortex core can be found
by replacing in the left hand side of Eq. (\ref{eqv})
$G_{{\bf nn'}}(\omega)$ by $G_0(\omega)\equiv 1/i\omega L_0+1/R_0$,
after which it is reduced to
\begin{equation}
-(1+i\omega L_0/R_0)\nabla^2 v=i\omega L_0 P(x,y) \;. \label{eqv2}
\end{equation}
The solution of Eq. (\ref{eqv2}) allows to find that away from the
core $v(x,y)$ decays as
\begin{eqnarray}
v(x,y) & \approx & \frac{1}{1+i\omega L_0/R_0} \label{vxy} \\
 & \times & \left[\frac{x}{4(x^2+y^2)}\ln(x^2+y^2)^{1/2}\right.
+\left.\frac{x(3y^2-x^2)}{16(x^2+y^2)^2}\right] \;,    \nonumber
\end{eqnarray}
which confirms the convergence of the lattice sum in Eq. (\ref{DL2}).

The frequency dependence of $(\Delta G_\Box)_2$, related with the
frequency dependent factor in Eq. (\ref{vxy}) is consistent with what
one expects from the driven oscillations of an overdamped particle in
effective harmonic potential. Quite naturally $(\Delta
L^{-1}_\Box)_2\equiv\lim_{\omega\rightarrow 0} i\omega(\Delta
G_\Box)_2(\omega)$ [as well as $(\Delta L^{-1}_\Box)_1$] does not
depend on the value of shunting resistance $R_0$.

\section{Results and discussion}

Thus we have found that at frequencies for which the diffusion of
vortices is effectively frozen
the main correction to the sheet inductance of a Josephson junction array
comes from the suppression of the superfluid density
rather then from the displacement of vortices in the effective
potential and contains logarithmic divergence.

In the presence of a small concentration of more or less uniformly
distributed vortices induced by the application of weak magnetic field
(with $f\ll 1$) the divergence in the expression for the vortex energy
is cut off at $r\sim f^{-1/2}$
(instead of at $r\sim N$), and the correction to $L^{-1}_\Box$ can be
rewritten as \begin{equation}
\Delta L^{-1}_\Box\approx - \frac{\pi}{4}L^{-1}_\Box f\ln\frac{1}{f}
\;\;.                                                \label{last}
\end{equation}
The same expression is also valid for other periodic lattices,
for example triangular.
When the screening effects related with the self-induced magnetic
fields of the currents in the array are taken into account,
the logarithmic factor in Eq. (\ref{last}) has to saturate
(with decrease of $f$) when the typical distance between vortices
becomes of the order of the magnetic penetration depth \cite{P,SK}
of the array.

Earlier the magnetoinductance of a Josephson junction array in the
regime of frozen vortex diffusion has been investigated for the
case of a fractal array with the structure of the Sierpinski
gasket \cite{KMM,Meyer}. With the help of the recursive
calculation using the self-similarity of the Sierpinski gasket it
has been found \cite{KMM} that the field dependence of the
correction to inductance in such system can be characterized by
the exponent $\nu_{L}=\ln(125/33)/\ln 4\approx 0.96$. On the other
hand, the application of a simplified approach \cite{Meyer}, which
in terms of this work is equivalent to consideration of only
$(\Delta L^{-1}_\Box)_1$, leads to a larger value $\nu_{E}=\ln
5/\ln 4\approx 1.16$. That means that in the case of a Sierpinski
gasket array the main contribution to the correction to sheet
inductance for small fields is coming from the term $(\Delta
G_\Box)_2$, which can be associated with oscillations of vortices.
The results of this work show that in periodic arrays the
situation is qualitatively different and the main contribution to
correction to inductance is coming from the suppression of the
superfluid density related with a non-uniform distribution of the
order parameter phases induced by the presence of vortices.

\section*{Acknowledgements}

The author is grateful to P. Martinoli for numerous interesting
discussions.
This work has been supported in part
by the Program "Quantum Macrophysics" of the Russian Academy of
Sciences, by the Program "Scientific Schools of the Russian
Federation" (grant No. 00-15-96747), by the
Swiss National Science Foundation and by the Netherlands Organization
for Scientific Research (NWO) in the framework of Russian-Dutch
Cooperation Program.

 \end{multicols}

\begin{thebibliography}{99}
\bibitem{FH}   A. T. Fiory and A. F. Hebard,
               AIP Conf. Proc. {\bf 58}, 293 (1980).
\bibitem{JGRLM}B. Jeanneret, J. L. Gavilano, G. A. Racine,
               Ch. Leemann and P. Martinoli,
               Appl. Phys. Lett. {\bf 55}, 2336 (1989).
\bibitem{HF}   A. F. Hebard and A. T. Fiory,
               Phys. Rev. Lett. {\bf 44}, 291 (1980);
               A. T. Fiory, A. F. Hebard and W. I. Glaberson,
               Phys. Rev. {\bf B} 28, 5075 (1983);
               A. T. Fiory, A. F. Hebard, P. M. Mankievich
               and R. E. Howard,
               Phys. Rev. Lett. {\bf 61}, 1419 (1988).
\bibitem{ML}   
               P. Martinoli and Ch. Leemann,
               J. Low  Temp. Phys. {\bf 118}, 699 (2000).
\bibitem{Ber}  V. L. Berezinskii,
               Zh. Eksp. Teor. Fiz. {\bf 59}, 907 (1970)
               [Sov. Phys.  JETP {\bf 32}, 493 (1971)];
               Zh. Eksp. Teor. Fiz. {\bf 61}, 1144 (1971)
               [Sov. Phys.  JETP {\bf 34}, 610 (1972)].
\bibitem{KT}   J. M. Kosterlitz and D. J. Thouless,
               J. Phys. C {\bf 5}, L124 (1972);
               J. Phys. C {\bf 6}, 1181 (1973);
               J. M. Kosterlitz,
               J. Phys. C {\bf 7}, 1046 (1974).
\bibitem{TJ}   S. Teitel and C. Jayaprakash,
               Phys. Rev. Lett. {\bf 51}, 1999 (1983);
               S. E. Korshunov,
               J. Stat. Phys. {\bf 43}, 17 (1986).
\bibitem{T}    R. Theron, J. B. Simond, J. L. Gavilano,
               Ch. Leemann and P. Martinoli,
               Physica B {\bf 165\&166}, 1641 (1990).
\bibitem{T1}   P. Martinoli, R. Theron, J.-B. Simond, R. Meyer,
               Y. Jaccard and Ch. Leemann,
               Physica Scripta {\bf T49}, 176 (1993).
\bibitem{T2}   R. Th\'{e}ron, J.-B. Simond, Ch. Leemann,
               H. Beck, P. Martinoli and P. Minnhagen,
               Phys. Rev. Lett. {\bf 71}, 1246 (1993).
\bibitem{T3}
               R. Th\'{e}ron, S. E. Korshunov, J. B. Simond,
               Ch. Leemann and P. Martinoli,
               Phys. Rev. Lett. {\bf 72}, 562 (1994).
\bibitem{M}    P. Martinoli, private communication.
\bibitem{LAT}  C. J. Lobb, D. W. Abraham and M. Tinkham,
               Phys. Rev. {\bf 27}, 150 (1983).
\bibitem{YS}   W. Yu and D. Stroud,
               Phys. Rev. B {\bf 50}, 13 632 (1994).
\bibitem{KMM}  S. E. Korshunov, R. Meyer and P. Martinoli,
               Phys. Rev. B {\bf 51}, 5914 (1995).
\bibitem{P}    J. Pearl, in {\em Low Temperature Physics - LT9},
               edited by J. D. Daunt, D. O. Edwards, F. J. Milford
               and M. Yacub (Plenum Press, New York, 1965), p. 566.
\bibitem{SK}   D. Stroud and S. Kivelson,
               Phys. Rev. B {\bf 35}, 3478 (1987).
\bibitem{Meyer}R. Meyer, J. L. Gavilano, B. Jeanneret, R. Th\'{e}ron,
               Ch. Leemann, H. Beck and P. Martinoli,
               Phys. Rev. Lett. {\bf 67}, 3022 (1991).

\end{thebibliography}
\end{document}